# Explainable Deep Learning for Tumor Dynamic Modeling and Overall Survival Prediction using Neural-ODE

*Mark Laurie[1,2], James Lu[1,*]*

(1) Modeling & Simulation/Clinical Pharmacology, Genentech, South San Francisco, CA, USA
(2) Department of Computer Science, Stanford University, Stanford, CA, USA
 * Corresponding author. Email: lu.james@gene.com. Address: Mail Stop 463A, 1 DNA Way, South San Francisco, CA 94080, USA

## Abstract

While tumor dynamic modeling has been widely applied to support the development of oncology drugs, there remains a need to increase predictivity, enable personalized therapy, and improve decision-making. We propose the use of Tumor Dynamic Neural-ODE (TDNODE) as a pharmacology-informed neural network to enable model discovery from longitudinal tumor size data. We show that TDNODE overcomes a key limitation of existing models in its ability to make unbiased predictions from truncated data. The encoder-decoder architecture is designed to express an underlying dynamical law which possesses the fundamental property of generalized homogeneity with respect to time. Thus, the modeling formalism enables the encoder output to be interpreted as kinetic rate metrics, with inverse time as the physical unit. We show that the generated metrics can be used to predict patients' overall survival (OS) with high accuracy. The proposed modeling formalism provides a principled way to integrate multimodal dynamical datasets in oncology disease modeling.

## Keywords

# 1. Introduction

The study of tumor growth dynamics has a long history, with early seminal efforts demonstrating the ability of mathematical models to describe experimental data [1][2]. Subsequently, there has been a plethora of mathematical models based on various frameworks (e.g. deterministic, stochastic, game theoretical, etc.) that have integrated aspects of the underlying biological processes (e.g. tumor heterogeneity and angiogenesis) and led to the generation of scientific insights [3]. For drug development applications, the modeling of tumor size dynamics from patient populations has become an important tool by which to characterize treatment efficacy. Currently, nonlinear mixed-effects modeling (NLME), based on structural models described with either algebraic or differential equations, is the most widely adopted methodology used by the pharmacometrics community [4]. Via modeling, the derived tumor dynamic metrics have found wide utility in supporting the development of anti-cancer drugs, ranging from early efforts [5] showing the ability of such metrics to predict Phase 3 overall survival (OS) from Phase 2 data, to broader applications [6][7] and use of the tumor dynamic metrics to predict the hazard ratio of clinical trials for a wide variety of solid tumor types [8]. While there have been prior efforts to use machine learning (ML) algorithms to map tumor metrics derived from NLME modeling [9] to OS, the derivation of the metrics themselves has not been attempted using ML. Although there has been much progress in *model-informed drug development* within oncology using tumor dynamic models, there remain many opportunities for additional applications, including personalized therapy [10].

A key area for the advancement of tumor dynamic modeling is increasing the ability to accurately predict future patient outcomes from early observed longitudinal data. If successful, this would increase the impact of predictive modeling for drug development and for personalized therapy. The paths for future progress may entail the utilization of high dimensional data that have become available through technological advances in the biomedical sciences (e.g. Digital Pathology [11], ctDNA [12] etc.), via the development of algorithms, or via a combination of both. Despite the many types of models that have been developed to support clinical decision making in oncology, there is increased appreciation that in order to effectively mine ever larger datasets, Artificial Intelligence (AI) approaches are required to complement existing statistical and mechanistic models [13].

While there are many neural network architectures that can be utilized to describe longitudinal data such as tumor size measurements, the formalism of Neural-Ordinary Differential Equations (Neural-ODE) [14][15] is an especially effective platform by which to combine the strengths of Deep Learning (DL), with the advantages of ODEs (which is amongst the most commonly used mathematical formalism in tumor dynamics modeling). The use of Neural-ODE in pharmacology has been successfully developed for pharmacokinetics (PK) [16] and pharmacodynamics (PD) [17] modeling, demonstrating promising predictivity results in the settings examined. In this work, we provide the foundational Tumor Dynamic Neural-ODE (TDNODE) modeling framework that consists of an encoder-decoder architecture. A key consideration in the methodological development is integrating the ML model with physical concepts [18], as this may enhance interpretability and enable making physically consistent predictions in temporal extrapolations beyond the training set [18]. While largely building upon the important work of [15] whereby a Recurrent Neural Network (RNN) encoder is used together with a Neural-ODE decoder, there are additional developments necessary to make it pharmacology-informed in terms of dynamical characterization of patients' tumor data that leverages the well-established oncology disease modeling framework [10]. In this work, we propose a principled way to normalize patients' tumor dynamics data in conjunction with scaling the encoder outputs in correspondence in order to maintain a specific *units-equivariance* [19] in the learned vector field. Thus, we not only enable the interpretation of the encoder outputs as tumor dynamic metrics (with the physical unit of inverse time), but also ensure a generalizable approach to predict patients' OS in a manner that rests upon the well-established tumor growth inhibition (TGI)-OS link. We show that by applying the proposed methodology together with our devised data augmentation approach, we can generate a model that makes unbiased predictions of future tumor sizes from early, truncated tumor size data. Furthermore, we show that the encoder-generated TDNODE metrics (that is, patient-specific kinetic parameters produced by feeding the longitudinal tumor data into the encoder) can predict patients' OS using the ML survival model XGBoost [21] with a performance that significantly surpasses the existing TGI-OS approach [21]. Finally, we show that the XGBoost model for OS can be interpreted using the Shapley Additive Explanation (SHAP) to quantify how the contribution of TDNODE metrics impacts tumor dynamic predictions [22] [23].

# 2. Results

## 2.1 Dynamical systems formulation enabling the interpretation of parameters as kinetic rates

We wish to discover the dynamical law governing patients' tumor dynamics using the Neural-ODE system shown below:

$$\frac{dz(t)}{dt} = f_\theta(z(t), p), \qquad t \in [0, T]. \qquad (1)$$

Here, $T$ is the final simulation time of interest, $f_\theta$ is a neural network parameterized by a set of weights $\theta$ to be learned across the patient population, $p \in \mathbb{R}^k$ represents the patient-specific kinetic parameters and $z(0) \in \mathbb{R}^c$ represents the patient-specific initial state of the system both of which are to be learned from individual patient data, and $z(\cdot): [0, T] \to \mathbb{R}^c$ represents the time-continuous solution being sought. The patient-specific parameters and initial state represent patient-to-patient variability obtained by passing the individual patient tumor data through the corresponding encoders. In particular, the current framework focuses on discovering the set of equations for tumor dynamics, independent of the choice of treatment and/or dosing.

In our formulation, we would like to be interpret $p$ as kinetic parameters with the physical unit of 1/[t] rather than simply an arbitrary abstract representation generated by a neural network with the sole aim of reproducing the tumor dynamic data. We do so by performing time-scaling operation on equation (1) and leveraging the notion of equivariance [24]: upon transformation to a dimensionless time $\hat{t} = t/T$ and using the chain rule, we have (please see the Appendix for the derivation):

$$\frac{dz(\hat{t})}{d\hat{t}} = f_\theta(z(\hat{t}), p) \times T, \qquad \hat{t} \in [0, 1]. \qquad (2)$$

We propose that in order to interpret $p$ as kinetic parameters, we should learn vector fields $f_\theta$ which satisfy the following generalized homogeneity [25] condition:

$$f_\theta(z, p) \times T = f_\theta(z, p \times T), \quad \forall z \in \mathbb{R}^c, p \in \mathbb{R}^k, T \in \mathbb{R}^+. \qquad (3)$$

Note that while vector fields $f_\theta(z, p)$ that are linear in $p$ enters would clearly satisfy (3), this condition does not equate linearity as there are nonlinear functions which satisfy (3) as well. From equations (2) and (3), it follows that:

$$\frac{dz(\hat{t})}{d\hat{t}} = f_\theta(z(\hat{t}), p \times T), \qquad \hat{t} \in [0, 1]. \qquad (4)$$

Equation (4) shows that in order for the dynamical system expressed in dimensionless time $\hat{t}$ to reproduce a given set dynamical data given in the original time $t$, its kinetic parameters need to scale in direct proportion to the corresponding time scaling factor: that is, $p \times T$. As $T$ can be an arbitrarily chosen positive real scaling factor, we introduce a patient-dependent data augmentation scheme such that a single temporal data trace (in original time) is truncated, and subsequently mapped to a number of different temporal traces expressed in rescaled time, $\hat{t}$. The augmented set of temporal traces under different rescaling is then used to train the model. In summary, we propose a data augmentation scheme involving various choices of time scaling factor $T$, thereby imbuing $p$ with the meaning of kinetic rate parameters. Refer to the <u>Methods</u> section for further details.

## 2.2 Model architecture generates longitudinal tumor predictions and kinetic parameters

We designed TDNODE with the intent to longitudinally predict tumor dynamics given an arbitrarily defined observation window, whereby the input data consists of tumor sum-of-longest-diameter (SLD) measurements and their respective times of measurement.

The main components of the TDNODE architecture consists of a set of two encoders that process sequential inputs and an ODE-solver decoder, as displayed in **Figure 1**: the *Initial Condition Encoder* consists of a recurrent neural network (RNN) that creates a 4-dimensional encoding which represents the initial condition of a given patient; the *Parameter Encoder* contains an attention-based Long Short-Term Memory (LSTM) flanked by laterally-connected linear layers, which produces a 2-dimensional representation of tumor kinetic parameters. The ODE-solver decoder uses the two encoder outputs as the initial condition and kinetic parameters respectively, to generate longitudinal SLD predictions. Following Chen *et al.* [14], the decoder consists of an ODE system whose vector field is represented as a neural network. In the ODE solver, numerical integration is carried out for the 4-dimensional state vector (which provides a latent representation of the time-varying tumor state) from the current time point to its corresponding values at the next requested time point, according to the dynamics provided by the vector field. Finally, a *Reducer* is used to reduce the state vector to a single number, so as to be compared with the measured SLD values. Refer to [Methods](Methods) for additional details.

## 2.3 TDNODE produces unbiased predictions of tumor dynamics

We examined the ability of TDNODE to reproduce and extrapolate tumor dynamics using the IMPower150 dataset [26]. As the aim of this work is to make longitudinal tumor predictions, we first excluded subjects from the dataset with less than or equal to one tumor size measurement. Subsequently, data were split randomly with respect to subject ID into a training set used to develop TDNODE (889 patients) and a test set used to assess TDNODE's tumor dynamic predictions (216 patients). In the training set, we performed the proposed method of augmentation that considers the subsampled subsets of collected patient time series data, which significantly increased the number of tumor dynamic profiles observed by TDNODE during training. Tumor SLD measurements were normalized with respect to the training cohort mean and standard deviation for both cohorts. Observation times were scaled with respect to the time of last measurement for each subject. To determine the last time of measurement, we deem all measurements made at a time $t$ less than or equal to the patient's observation window ($w_i$) since the start of treatment as *seen*, measurements that TDNODE is able to observe and *estimate* in its tumor dynamic prediction curve. All measurements collected at time $t$ greater than the $w_i$-week observation window are conversely deemed as *unseen*; here TDNODE is assessed in its ability to *extrapolate* or infer tumor dynamics using only data collected from within its observation window. In this study, we let $w_i = 32$ weeks represent the observation window for each subject.

We show in **Table 1** that TDNODE can accurately extrapolate tumor dynamics on the test set for $t > w_i$ across treatment arms in the IMPower150 dataset. Here the bootstrapped median root-mean-squared error (RMSE) and $R^2$ score on the extrapolation component of the test set were shown to be 9.69 and 0.88, respectively. We note that these data were never seen by TDNODE during model development and inference.

Training set RMSE and $R^2$ performance can be seen in **Supplementary Table 4**. Here TDNODE achieved a bootstrapped median $R^2$ score of 0.95 on all measurements on both the training and test sets.

TDNODE was also shown to produce unbiased predictions, as shown in **Figure 2** for both the training and test sets. In particular, TDNODE predictions did not exhibit systematic trends with respect to the measured SLD values (**Figures 2a-2b**). Moreover, TDNODE exhibited no systematic bias in a statistically significant manner in the tumor size predictions with respect to the time of measurement (**Figure 2c-2d**). These findings demonstrate promise for TDNODE to both longitudinally model and forecast tumor size data, in contrast to classical TGI models which were shown to carry systematic biases of tumor size predictions with respect to observation time [20]. Refer to **Supplementary Figures 1 and 2** for plots showing these comparisons for each treatment arm in the IMPower150 dataset.

We illustrated TDNODE's predictive abilities using selected patient longitudinal SLD measurements from the test set. To this end, we adjust $w_i$ to be 16, 24, or 32 weeks for all subjects. We observe in **Supplementary Figure 3** how the TDNODE predicted tumor profiles for these patients change with respect to increases in observation window. We also observe in **Supplementary Table 5** how TDNODE's extrapolation performance improves as the observation window is increased. Taken together, our results show that TDNODE can make qualitatively appropriate continuous predictions of tumor dynamics at the subject level and that TDNODE makes more accurate predictions as the observation window is increased, as expected. Additional patient longitudinal tumor dynamic predictions can be seen in **Supplementary Figure 4**.

## 2.4 TDNODE enables superior prediction of Overall-Survival compared to existing TGI-OS models

The parameter encoder module of TDNODE produces a 2-dimensional encoding that can be used to predict patients' OS. In a similar manner to using TGI metrics to predict OS [9][27], we used this output as the input to an XGBoost ML modeling approach used previously in [27] (see parameters in **Supplementary Table 3**). We used the same sets of training-test patient split to evaluate OS predictions. Only encoder outputs from the training set patients were used to construct the OS ML model. This model did not observe test set encoder metrics until after the OS ML model was developed. We evaluated OS predictivity using the c-index [28], which was calculated using the OS ML model's output and the OS status of each patient from the clinical trial. On the training set, we evaluated OS predictivity by 5-fold cross-validation with random splitting of the data.

We compared this statistic to that obtained using existing set of all TGI metrics (consisting of tumor growth rate (KG), tumor shrinkage rate (KS) and the time to tumor growth (TTG) [6]), which were obtained using a non-linear mixed-effects (NLME) model fitted to data from all patients, from which the individual parameters were obtained [8], as is standard practice in pharmacometrics modeling. For both models, we evaluated OS predictivity using just the generated metrics, or in conjunction with eleven baseline covariates (**Supplementary Table 2**) obtained at the start of each subject's enrollment in the clinical trial [27]. **Figure 3** displays how the ML-predicted survival probabilities with 95% confidence intervals for each treatment arm align with that of the Kaplan-Meier (KM) curves of the patients in the test set.

The results shown in **Table 2** demonstrate that the ML model for predicting OS based on TDNODE metrics (referred to as TDNODE-OS.ML) resulted in significantly increased predictive performance as compared to TGI metrics (referred to TGI-OS.ML) and is true using either the metrics alone, or using the metrics in conjunction with the previously used eleven baseline covariates. Furthermore, in the case of OS prediction using TDNODE derived metrics, we observe no significant loss in performance in OS predictivity between the validation set patients and the test set patients by comparing the c-index values. Furthermore, the predictive performance levels of TDNODE-OS.ML showed marginal improvements after incorporating the eleven baseline covariates. The result suggests that the TDNODE metrics extract sufficient information from the longitudinal tumor size data such that the additional eleven baseline covariates no longer provided additional predictive value. This is in contrast to the case of TGI-OS.ML, where utilizing the eleven baseline covariates resulted in a sizable improvement in OS predictivity.

## 2.5 TDNODE metrics can explain patient Overall Survival

To enhance the interpretability of TDNODE, we designed the formulation such that a single dynamical law is learned from the tumor dynamics data with patient-to-patient variability explained by kinetic parameters expressed with the units of inverse time. In addition to showing that the TDNODE-generated metrics can be used to predict OS, here we demonstrated that the relationship between the metrics and OS can be explained at a qualitative level as can be seen in **Figure 4**. Using principal component analysis (PCA) [29], we obtained the first principal component from the set of 2-dimensional encoder metrics and showed that it exhibits similar OS predictability to that of the TDNODE-generated metrics when using XGBoost, achieving a c-index of 0.82 on the training set and 0.81 on the test set (see **Supplementary Table 6** for details). Shapley Additive Explanation (SHAP) [22] [23] analysis of this XGBoost model shows a positive impact upon the OS expected survival time with respect to increases in this axis.

Given this finding, we evaluated the impact of perturbations of this component on the tumor dynamics of individual patients. In **Figure 4**, we find that increases in this component yield monotonic decreases in the longitudinal tumor dynamic predictions for all patients in the training and test sets. This finding aligns with that of the SHAP analysis that we performed, as increases in tumor sizes are expected to be associated with decrease in OS. SHAP summary plots of the XGBoost ML model using the 2-dimensional encoder output of TDNODE can be found in **Supplementary Figure 5** and **Supplementary Figure 6**, and additional examples of feature dependence plots for randomly selected patients can be found in **Supplementary Figure 7** and **Supplementary Figure 8**.

## 3. Discussion

We presented a deep learning methodology to discover a predictive tumor dynamic model from longitudinal clinical data. In essence, the methodology leverages neural-ODEs [14][15] formulated in such a way that a single underlying dynamical law (represented by the vector field) is to be discovered from the patient population data, with patient-to-patient variability in the tumor dynamics data explained by differences in both the individual initial state as well their kinetic rate parameters. Furthermore, we introduced an equivariance property in the vector field under time rescaling transformation, so as to enable the interpretation of the learned patient embedding as kinetic parameters, with the units of inverse time. By estimating the individual patients' kinetic parameter values (or metrics) from the longitudinal tumor size data, they can then be used to predict the patients' OS using a ML model. Our proposed use of leveraging longitudinal tumor size data to generate metrics and subsequently OS predictions, follows the well-established TGI-OS paradigm [5] [8] [10] which has been widely applied to support oncology drug development. On the other hand, prior TGI-OS approaches have all relied on the human modellers choosing appropriate parametric functions for both the tumor dynamics equations as well as the statistical models for survival. In this work, we endeavor to leverage machine intelligence to discover the underlying models while retaining a one-to-one correspondence with the traditional TGI-OS paradigm on a conceptual level. Compared to many other alternate approaches for modeling time series data, our proposed TDNODE methodology has the benefit of explainability brought about by the following: (1) the bottleneck of the encoder-decoder architecture enables accounting for patient-to-patient variability in a parsimonious manner; (2) the equivariance condition of the vector field enables interpretation of patient embeddings as kinetic parameters; (3) the dynamical system nature of the formulation enables a direct connection between salient aspects of tumor trajectories and their impact on the predicted OS.

The proposed methodology was applied to data from IMPower150, a phase 3 clinical trial of NSCLC patients. We showed on this dataset that the proposed TDNODE methodology overcame a key obstacle of the current TGI modeling approach, namely its ability to predict in an unbiased manner the future tumor sizes from early longitudinal data. These results can be explained by the formulation of TDNODE in minimizing a loss function and the proposed data augmentation scheme of feeding the early tumor dynamics data into the encoder while ensuring accurate extrapolations for the future dynamics. In contrast, the approach of population modeling based on nonlinear mixed effects [30] aims to characterize model parameters at both the population and individual levels rather than aiming to minimize errors in model predictions over an unseen horizon. Another benefit of utilizing the deep learning approach was the discovery of tumor dynamics metrics, beyond the tumor growth and shrinkage rates (i.e., KG and KS respectively) that have been widely used in TGI literature [8] [10]. We showed that the TDNODE metrics obtained were better predictive of patient OS than the existing ones, as demonstrated by the sizably higher c-indices achieved.

There remain several areas for further research. While the results shown here demonstrate significant promise in the setting of NSCLC patients, it remains to be applied to other solid and hematological cancer types. Of note, as TGI models of the same structure [3] are often applied across different tumor types, it may be advantageous to apply TDNODE to identify the best set of pan-tumor dynamical equations. Additionally, our current TDNODE model does not incorporate dosing or PK; however, such extensions are possible areas for further work. In order to further improve the TNODE predictive performance, hyperparameter optimization and the experimentation of alternate numerical solvers (e.g., [41]) are promising avenues. While this work sets the mathematical foundations for longitudinal tumor data, further expansions to incorporate multimodal, high dimensional data [31] of both static and longitudinal nature remain active areas for future development.

# 4. Methods

## 4.1 Data summary

Longitudinal tumor sum-of-longest-diameters (SLD) data were collected from the IMpower150 clinical trial in which 1,184 chemotherapy-naive patients with stage IV non-squamous non–small cell lung cancer (NSCLC) were enrolled [32]. This Phase 3, randomized, open-label study evaluated the safety and efficacy of atezolizumab (an engineered anti-programmed death-ligand PD-L1 antibody) in combination with carboplatin + paclitaxel with or without bevacizumab compared to carboplatin + paclitaxel + bevacizumab [32]. All patients provided written consent prior to enrollment. Patients assigned to Arm 1 were administered atezolizumab + carboplatin + paclitaxel (*n = 400*). Patients assigned to Arm 2 were administered atezolizumab + carboplatin + paclitaxel + bevacizumab (*n = 392*). Patients assigned to Arm 3 were administered carboplatin + paclitaxel + bevacizumab (*n = 392*). In Arms 1 and 2, atezolizumab was administered as an IV infusion at a dose of 1,200mg Q3W until a loss of clinical benefit was observed. In all arms, carboplatin was administered at 6mg /mL min$^{-1}$ Q3W for 4 cycles, 6 cycles, or until loss of clinical benefit, whichever came first. In Arms 2 and 3, bevacizumab was administered as an IV infusion at a dose of 15 mg kg$^{-1}$ Q3W until disease progression, unacceptable toxicity, or death. In all arms, paclitaxel was administered as an IV infusion at a dose of 200mg m$^{-2}$ Q3W for 4 cycles, 6 cycles, or until loss of clinical benefit, whichever came first. Full details of study design can be found in [26].

## 4.2 Data processing

### 4.2.1 Data definition

Let $\psi = \{y_1, y_2, \cdots, y_n\}$ represent the set of SLD measurements from $n$ subjects in the dataset, where $y_i = (y_{i,1}, y_{i,2}, \cdots, y_{i,m_i})$ is a list of SLD measurements (typically measured in mm) sorted by time, whereby $y_{i,j}$ corresponds to the $j$th SLD measurement for subject $i$. Here $m_i$ corresponds to the number of SLD measurements for subject $i$. Similarly, let $\Gamma = \{\tau_1, \tau_2, \cdots, \tau_n\}$ represent the corresponding observation times of $n$ subjects in the dataset, where $\tau_i = (\tau_{i,1}, \tau_{i,2}, \cdots, \tau_{i,m_i})$ is a sorted list of $m_i$ observation times (typically expressed in weeks), such that $\tau_{i,j}$ corresponds to the $j$th observation time for subject $i$. We exclude patients with less than or equal to one post-treatment observation since longitudinal modeling would not be applicable. Using these definitions and the IMPower150 dataset, we have a total number of 1,105 eligible subjects.

### 4.2.2 Data splitting

We then randomly allocate 80% of subjects ($n_{train} = 889$) into the training set and 20% of subjects ($n_{test} = 216$) into the test set (consisting of 73, 68 and 75 patients in treatment arms 1, 2 and 3 respectively). Here $\Psi_{train}$ and $\Psi_{test}$ represent the SLD measurements of subjects in the training set and test set, respectively. Similarly, $\Gamma_{train}$ and $\Gamma_{test}$ represent the observation times of subjects in the training and test set, respectively.

### 4.2.3 Definition of pre-treatment and post-treatment measurements

The dataset contains tumor SLD measurements collected across two distinct segments: 1) pre-treatment and 2) post-treatment. We define the pre-treatment measurements as SLD measurements collected prior to the initiation of active treatment. Conversely, we define post-treatment measurements as SLD measurements collected after the active treatment has been initiated. As the time for the start of treatment is taken to be 0, $y_{i,j}$ is considered pre-treatment when $\tau_{i,j} < 0$, and post-treatment when $\tau_{i,j} \geq 0$.

### 4.2.4 Definition of Observation Window

We also defined a set of observation windows for subjects in both training and test cohorts. Here, an observation window is defined as the quantity of post-treatment time in which a subject has been observed. Let $W = \{w_1, w_2, \cdots, w_n\}$ represent the set of observation windows for all $n$ subjects such that $w_i$ represents the observation window for subject $i$. This scalar corresponds to the amount of *seen* data that TDNODE uses to predict tumor dynamics and generate parameter encodings. Thus, any measurements outside the observation window are deemed as *unseen* by TDNODE and subsequently not used as input. In this study we let $w_i = 32$ weeks for all patients.

From this definition, consider patient $i$ and let $l(i) = argmax_j(\tau_{i,j} \leq w_i)$ represent indices for measurement such that $\tau_{i,l(i)}$ and $y_{i,l(i)}$ correspond to the last observed time of measurement and last observed SLD value; that is, the last observed SLD value to be used as input for TDNODE. Hence, for patient $i$, all (integer) indices $j$ between 1 and $l(i)$ represent observed measurements, and all indices greater than $l(i)$ represent measurements unseen by the model in either training or testing.

### 4.2.5 Normalization of *SLD*

We compute the mean and standard deviation of all SLD values within the training set. In this regard, let $\mu$ represent the arithmetic mean of $y_{i,j}$ computed over all $i, j \in \Psi_{train}$, and let $\sigma$ represent the standard deviation of $y_{i,j}$ computed over all $i, j$. Then, we perform Z-score normalization on $y_{i,j}$ using $\mu$ and $\sigma$:

$$\tilde{y}_{i,j} = \frac{y_{i,j} - \mu}{\sigma} \quad (5)$$

where $\tilde{y}_{i,j}$ is the Z-score normalized $j$'th measurement of subject $i$. Here we let $\tilde{\Psi}_{train}$ and $\tilde{\Psi}_{test}$ represent the normalized SLD values of all patients in the training and test sets, respectively.

### 4.2.6 Normalization of *time*

We introduce a patient-specific method to normalize the observation times of each subject in both the training set and test set. For each patient, we declare the last observed time of measurement as a scaling factor that is used on that patient's time series. We then use the last observed measurement time to normalize its respective list of observation times $\tau_i$ for all $i, j$:

$$\tilde{\tau}_{i,j}(\tau_i, l(i)) = \frac{\tau_{i,j}}{max_{j \leq l(i)} \tau_{i,j}}, \quad (6)$$

such that $\tilde{\tau}_{i,j} \leq 1$ up to the observation window and $\tilde{\Gamma}_{train}$ and $\tilde{\Gamma}_{test}$ contain the normalized observation times for all subjects.

### 4.2.7 Augmentation

We wish for TDNODE to be generalizable such that it can longitudinally model a representative set of tumor dynamic profiles commonly seen in clinical trials. Doing so requires that the training set be large and diverse, containing a wide variety of tumor dynamic profiles upon which TDNODE can generalize when modeling tumor dynamics on unseen data. To increase the diversity of tumor dynamic profiles in the training set, we introduce a new subsampling strategy that we apply to training set observations $\Gamma_{train}$ and $\Psi_{train}$. Recall that $l(i)$ is an index that corresponds to the last observed measurement for subject $i$. Hence, $\tilde{\tau}_{i,1:l(i)}$ and $\tilde{y}_{i,1:l(i)}$ represent the normalized post-treatment tumor dynamics for patient $i$, and components of these lists are used to generate subject $i$'s parameter encoding. We took $l(i)$ to truncate each subject's observed tumor dynamics corresponding to the respective observation window, $w_i$.

For each patient $i$ in the training set, data augmentation is performed in the following manner: we consider the set of all integer intervals ranging from one to the number of measurements for the subject $m_i$, that is $\widehat{j(\iota)} \equiv \{[1,j] | 2 \leq j \leq m_i\}$, and in each case derive the associated sets of normalized SLD and time values: $\{\tilde{y}_{i,j} | j \in \widehat{j(\iota)}\}$ and $\{\tilde{\tau}_{i,j}(\tau_i, \max(j)) | j \in \widehat{j(\iota)}\}$.

The result of this subsampling method is an additional 4,671 tumor dynamic profiles to be used during training, which significantly boosts the size of the training set to a total of 5,560 subsampled patients.

## 4.3 Formulation of the patient-dependent initial condition

Here we describe the formulation of the representation of subject $i$'s initial tumor state:

$$z_i(0) = g_\theta(\tilde{\tau}_{i,1:\beta_i}, \tilde{y}_{i,1:\beta_i}) \quad (7)$$

where $z_i(0) \in \mathbb{R}^c$ represents subject $i$'s $c$-dimensional initial tumor state, and $g_\theta$ is the initial condition encoder neural network parametrized by $\theta$, that takes as input subject $i$'s pre-treatment tumor size measurements ($\tilde{y}_{i,1:\beta_i}$) and observations times ($\tilde{\tau}_{i,1:\beta_i}$) to produce $z_i(0)$. Here $\beta_i$ is an index that corresponds to the last pre-treatment measurement for subject $i$:

$$\beta_i = \max(argmax_j(\tau_{i,j} \leq 0), 1) \quad (8)$$

where all indices less than or equal to $\beta_i$ correspond to pre-treatment measurements, and all indices greater than $\beta_i$ correspond to post-treatment measurements. If a subject has no pre-treatment measurements, $\beta_i$ is set to 1, indicating that the first measurement represents the pre-treatment tumor size for subject $i$. A description of the initial condition encoder's architecture can be found in the Model Architecture section and for further information please refer to the model code provided at the locations given in **Supplementary Note 1**.

## 4.4 Implementation of the patient-dependent parameter encoding and temporal rescaling

As discussed in the Results section, we leverage equation (4) to transform data using a subject dependent temporal rescaling into a unit time interval. This is implemented in the following manner: the kinetic parameter encoding for subject $i$ is computed as,

$$p_i = h_\theta(\tilde{\tau}_{i,\beta_i:l(i)}, \tilde{y}_{i,\beta_i:l(i)}) \cdot \tau_{i,l(i)} \quad (9)$$

where $p_i \in \mathbb{R}^k$ represents subject $i$'s $k$-dimensional representation of tumor kinetic parameters and $h_\theta$ is the TDNODE parameter encoder neural network, parametrized by $\theta$, that takes as input subject $i$'s post-treatment observed tumor dynamics to produce $p_i$. We wish to reference only post-treatment tumor dynamics, starting from the last pre-treatment measurement. To do so, we use the value at index $\beta_i$ to reference the last pre-treatment measurement for subject $i$. Additionally, since we assume the system evolves autonomously from time zero, we set $\tilde{\tau}_{i,\beta_i} = 0$. Note that in equation (9) the last observed time $\tau_{i,l(i)}$ serves as the temporal scaling $T$ of equation (4). A description of this network's model architecture used to generate $p_i$ can be found in the Model Architecture section and for further information please refer to the model code provided at the locations given in **Supplementary Note 1**.

## 4.5 Neural-ODE solution

We solve the Neural ODE in equation (4) using the initial condition encoding $z_i(0)$ and the parameter encoding $p_i$. We numerically integrated this ODE system using $f_\theta$, beginning at $\tau = 0$ and ends at $\tau = \tilde{\tau}_{i,m_i}$ (the normalized last measurement time for subject $i$). If batched solving is enabled, the solving process continues until the normalized last measurement time for the subject in the batch with the highest normalized last measurement time. Subsequent to the generation of the $c$-dimensional tumor state evolution, it is passed to a reducer which results in a series of scalar tumor size measurements. The ODE solver used was Dormand–Prince 5(4) (`dopri5` as implemented within the `torchdiffeq` library [35]) and the continuous adjoint solution was used for backpropagation; please see the model code provided at the locations given in **Supplementary Note 1** for further information.

## 4.6 Model architecture

TDNODE consists of four modules: an initial condition encoder that transforms pre-treatment tumor size data into a $c$-dimensional representation of the subject's initial tumor state, a parameter encoder that transforms post-treatment tumor size data into a $k$-dimensional representation of the subject's observed tumor dynamics, a neural ODE decoder module that computes a continuous series of $c$-dimensional tumor state representations, and a reducer module that produces the final series of continuous scalar predictions. Refer to **Figure 1** for a visual representation of the computational graph that utilizes these modules.

### 4.6.1 Initial condition encoder

The initial condition encoder, denoted as $g_\theta$, takes as input a tensor representing the pre-treatment measurements and times of measurement and generates a batch of $c$-dimensional representation of each subjects' pre-treatment tumor state $z_i(0)$. Here the input is of shape $B \times M_{\beta,max} \times 2$, where $B$ is the specified batch size, or number of subjects to process and $M_{\beta,max}$ is the number of pre-treatment measurements for the subject with the most pre-treatment measurements in the batch. The initial condition encoder consists of a multi-layer gated recurrent unit (GRU) recurrent neural network (RNN); the output of the RNN is processed by a single fully connected layer, producing a tensor with shape $B \times c$.

### 4.6.2 Parameter encoder

The parameter encoder, denoted as $h_\theta$, takes as input a tensor representation the truncated post-treatment measurements and times of measurement and generates a $k$-dimensional representation of each subject's post-treatment tumor dynamics up to an arbitrarily defined observation window. The parameter encoder takes as input a tensor of shape and $B \times (M_{s,max} - 1) \times 4$ produces a batch of parameter encodings $p_i$ of shape $B \times k$; here $M_{s,max}$ denotes the number of *seen* measurements for the subject with the highest number of observed measurements in the batch. Finally, we apply equation (9) to obtain a batch of patient-normalized encodings $p_i$ with the same shape. The parameter encoder's architecture utilizes multi-headed attention in as well as with residual fully connected layers. The multi-headed attention layer requires as input a key, value, and query. Here the key and value are generated by separate fully connected layers. The query is generated using a deep residual neural network and a Long Short-Term Memory (LSTM) network with 100 hidden units. The outputs of the query and attention layer are subsequently concatenated. Finally, a deep residual network consisting of fully connected layers is used to generate the patient-specific kinetic parameters. Please refer to the model code referred to in **Supplementary Note 1** for further details.

### 4.6.3 Neural ODE vector field

Upon creation of the batch of parameter encodings $p_i$ and initial conditions $z_i(0)$, we use equation (1) to solve the neural ODE system and generate a solution $z(\cdot): [0, T] \to \mathbb{R}^c$ with shape $B \times q \times c$, where $q$ is the number of measurements to be obtained in the interval 0 to $T$. Here $T$ is the upper bound of time value in numerical integration, which is equivalent to the maximum of the last times of measurement for subjects in the batch. The numerical integration is carried out with a neural network decoder, $f_\theta$, which takes as input the batch of initial conditions $z_i(0)$ and parameter encodings $p_i$ to produce the time derivatives used to compute the next $c$-dimensional tumor state. This tumor state is then used as input with the parameter encoding to solve to compute the time derivative to compute the next state, and so forth. We used the continuous adjoint solving method in this work. Here we define $f_\theta$ as a series of fully connected layers with residual connections. Each series of fully connected layers is interspersed with SELU activation functions [33]. The resulting solution $z(\cdot)$ is then converted back into a scalar space by a Reducer module, described below. For further details, please refer to the model code at the locations provided in the **Supplementary Note 1**.

### 4.6.4 Reducer

The generated batch of solutions to the neural ODE system $z(\cdot)$ is represented as a batch of $B$ set of $c$-dimensional tumor states that required conversion to a series of scalar SLD predictions. Here we instantiate a simple neural network reducer that takes the $c$-dimensional batch of solutions produced by the NODE decoder and converts it into a batch of $B$ scalar SLD predictions that represent the predicted tumor sizes for each patient. We implemented the Reducer as a series of fully connected layers interspersed with SELU activation functions.

## 4.7 Model Development

### 4.7.1 Instantiation

In this study, we set the initial condition encoder output dimension ($c$) to 4. Here we set the hidden dimension in the GRU to 10. As the input to the initial condition encoder is a tensor of time-observation pairs, we set its input dimension to 2.

For the parameter encoder, we set the output dimension ($k$) to 2. As the input is a tensor of partitioned time-observation pairs, we set the parameter encoder's input dimension to 4. We set the dimensionalities of all networks in the preprocessor encoder network to 4 and use a single-headed self-attention mechanism with output dimensionality set to 100. The LSTM module's output is also set to 100. These outputs are concatenated and used as input to the post-processor modules, which compresses this representation into the 2-dimensional encoding output.

The NODE decoder module takes as input a 6-dimensional tensor consisting of the 4-dimensional initial condition encoding and 2-dimensional parameter encoding. Each fully connected layer has a hidden dimension of 21 and produces a 6-dimensional representation of the tumor state. Since the last 2 dimensions of this output correspond to that of the parameter encoder, we set these values to 0 to signal that the parameter encoding remains constant throughout the solving process. The result is a series of 6-dimension solutions at every time point requested.

Finally, the reducer takes as input the set of 4-dimensional solutions at each step to produce a 1-dimensional series of SLD predictions for each patient using a series of fully connected layers. Note that, since the output of the NODE module is a set of 6-dimensional encodings at every time step, the reducer module uses only the first $c = 4$ values of the obtained solution.

### 4.7.2 Partitioning

Equations (4) and (6) use truncated tumor dynamic profiles to generate initial condition parameter encodings for each patient, respectively. The input to equation (4) is a tensor of pre-treatment measurements, where each row corresponds to a single observation and its corresponding time of measurement. Conversely, the input to equation (6) is a tensor of *partitioned* post-treatment observed measurements. Here each row corresponds to a pair of adjacent observations. For instance, for a patient with 4 time series measurements, of which 3 are deemed as seen, a partitioned tensor of shape 2 x 4 is created. In the first row, the first and second measurements are concatenated. In the second row, the second and third measurements are concatenated. Each row has length 4 as each pair of measurements consists of a time and SLD observation. We implement this partitioning mechanism to better enable learning of each patient's post-tumor state.

### 4.7.3 Batching

To enable higher throughput training, we propose a batching operation to simultaneously generate solutions for multiple patients at a time. Because each patient may have a different number of measurements, we produce masks that screen each patient solution for the predictions that correspond to observed data. For each batch, the union of times is collected as a single array. Labels are also concatenated together, with the position of each label corresponding to the appropriate index in the time tensor. A mask tensor with the same shape is generated as well, with 1s representing valid positions and 0s representing positions to exclude. Pre-treatment and post-treatment tensors of each patient in the batch are stacked. Left-padding of variable length is applied, with the pad value equivalent to the first time-SLD observation pair. IDs and cutoff indices for each are also stacked and used in each iteration.

### 4.7.4 Loss Calculation

During model development, batched tensors of shape $B \times L$ representing the continuous time series SLD predictions of $B$ patients are produced during each iteration. This prediction tensor is utilized in conjunction with a label and mask tensor (each of the same shape) to calculate the RMSE for the iteration used to adjust TDNODE's weights via backpropagation. In this loss function, the mask, a one-hot tensor with 1's representing the locations in which to tabulate the loss, is used to parse the prediction tensor for predictions with times that correspond to actual observed SLD measurements. After this filtering is applied, the RMSE is calculated using the labels tensor. Backpropagation is then carried out to optimize the weights of all four TDNODE modules.

### 4.7.5 Hyperparameter configuration

We trained TDNODE for 150 epochs using ADAM optimization [34], an L2 weight decay of 1e-3, a learning rate of 5e-5, an ODE tolerance of 1e-4, a batch size of 8, and an observation window of 32 weeks. Refer to **Supplementary Table 1** for additional information on the hyperparameter configuration of TDNODE.

### 4.7.6 Libraries used

In this implementation, we used the `torchdiffeq` library to carry out the solving process [35]. Other notable data science libraries include `pandas`, `numpy`, and `scipy` [36], [37]. We used the `Pytorch` deep learning framework to develop and evaluate TDNODE [38]. Survival analysis and SHAP analysis were performed using the `lifelines` and `shap` libraries, respectively [23], [39]. We used `matplotlib` to generate the majority of plots in this study [37]. Refer to the `environment.yml` file in the model code (see locations provided in **Supplementary Note 1**) for additional packages used.

## 4.8 Model Evaluation & Analysis

### 4.8.1 TDNODE Benchmarking

We used root-mean-squared error (RMSE) and $R^2$ score to evaluate performance of TDNODE on the training and test sets. To obtain bootstrapped RMSE and $R^2$ scores, we obtained the median and standard deviation of 1000-sample RMSE and $R^2$ distributions, where each sample RMSE and $R^2$ value is calculated from *n* prediction-label pairs sampled with replacement from the dataset. Refer to **Table 1** and **Figure 2** for additional performance metrics of TDNODE with respect to treatment arm and dataset and scatter plots visualizing the association between TDNODE-generated predictions and observed SLD data with respect to treatment arm and dataset.

### 4.8.2 Residual versus time analysis

Residuals between the TDNODE predictions at time points with observed measurements and their corresponding were obtained for all patients in the training and test sets (**Figure 2**). A bootstrapped LOWESS curve with 95% confidence interval (CI) was applied to assess if there was any systematic bias in the predictions with respect to time. Refer to **Supplementary Figure 2** for a visualization of residual versus time plots with respect to each treatment arm and dataset.

### 4.8.3 Generation of patient-level SLD predictions

For all patients in the training and test sets, we generated continuous longitudinal SLD predictions (**Figure 3**) with observation windows set to 16 weeks, 24 weeks, and 32 weeks. Refer to Model Architecture and Neural ODE Solving Mechanism for details on how these continuous predictions were produced.

### 4.8.4 Generation of TDNODE derived metrics for each patient

The trained parameter encoder component of TDNODE was used to generate the 2-dimensional kinetic rate metrics for each patient. These two variables were used in downstream Principal Component Analysis (PCA), XGBoost-ML for OS hazard rate prediction, and patient-level plots that assess the effect of directed perturbations of these variables.

### 4.8.5 Principal component analysis

We performed principal component analysis (PCA) [29] on the 2-dimensional kinetic rate metrics produced by TDNODE for each patient. Refer to **Figure 4** for a description on how these principal axes were obtained and how they were used to link tumor dynamics and OS.

### 4.8.6 Prediction of OS using XGBoost-ML from TDNODE-generated metrics

We used the first principal component to predict OS for all patients in the dataset using XGBoost-ML [21][40][23]. In the training set, we obtained c-indices representing the OS prediction accuracy using 5-fold cross-validation and reported the median c-index. We then used the entire training set to train an XGBoost-ML model tasked to predict OS on all patients in the test set, which was used to generate hazard rates for patients in the test set. Refer to **Supplementary Table 3** for XGBoost-ML training parameters and the observed c-indices with respect to input parameters used. Here we used the following six different sets of input variables: TDNODE-generated parameter encodings with and without baseline covariates; the two principal component values for each patient with and without baseline covariates; the first principal component value with and without baseline covariates.

### 4.8.7 SHAP

We used SHapley Additive exPlanations (SHAP) to identify the degree and contributions of input variables to our input models[23]. We performed SHAP analysis on the XGBoost-ML model using just the first principal component obtained from the TDNODE-generated metrics to identify its directionality. We also performed SHAP analysis on the XGBoost-ML model that directly utilized the TDNODE-generated metrics and on the XGBoost-ML model that utilized the metrics in conjunction with the eleven baseline covariates obtained for each patient. Refer to **Supplementary Figure 5** and **Supplementary Figure 6** to view the SHAP summary plots for relevant variables in each of these models.

### 4.8.9 Generation of subject feature dependence plots

The first principal axis was used to systematically perturb the 2-dimensional parameter encodings produced for each patient. This component's direction was transformed into the data space, yielding the unit vector $[0.701, 0.701]$. For each patient, we plotted the original predicted dynamics, along with $500$ predicted dynamics with systematically perturbed encodings. The magnitude of perturbation was defined with range $[-2, 2]$. Refer to **Supplementary Figure 7** and **Supplementary Figure 8** for additional examples of feature dependence for selected patients in the training and test set.

### 4.8.10 Generation of OS survival curves

We used the XGBoost-ML model that utilized only the first principal component of the TDNODE-generated encodings to model the expected survival times of each patient in the test set with respect to the treatment arm. Refer to **Figure 3** to view examples of these plots.

## Code availability

The model code is available at: https://github.com/jameslu01/TDNODE

## Acknowledgement

We would like to acknowledge Kenta Yoshida, Ji Won Park, Dan Lu, Rene Bruno, Chunze Li, Jin Y. Jin and Amita Joshi for their input and feedback on this work. No funding was received for this work.

# Appendix

The full derivation of equation (2) is as follows:

Let $t(\hat{t}) = \hat{t} \times T$ and $y(\hat{t}) = z(t(\hat{t}))$. Then, we have:

$$\frac{dy(\hat{t})}{d\hat{t}} = \frac{dz(t(\hat{t}))}{dt} \times \frac{dt(\hat{t})}{d\hat{t}} = f_\theta(z(t(\hat{t})), p) \times T = f_\theta(y(\hat{t}), p) \times T.$$

Thus, we obtain equation (2) by renaming $z(\hat{t}) = y(\hat{t})$.

# Figures

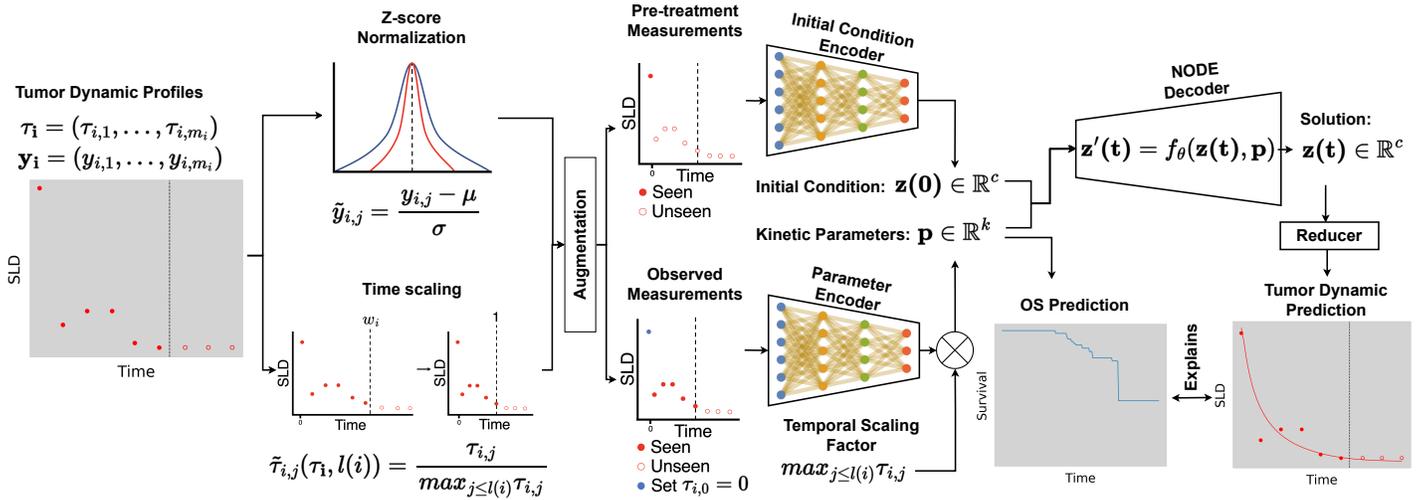

*Figure 1: Schematic representation of tumor dynamic neural-ODE (TDNODE).* *The deep learning model is designed to discover the underlying dynamical law from tumor dynamics data and use the identified kinetic parameters to predict patient Overall Survival (OS). Time series tumor dynamic data were split at the patient level into a training and test set. Times of last observation were obtained and the augmented patient time series data were created. The SLD data were Z-score normalized using the mean and standard deviation from the training set; the measurement times were scaled at the patient level using each patient's last observed measurement time. Pre-treatment and truncated post-treatment tumor dynamic profiles were fed into the initial condition and parameter encoders of TDNODE, respectively. Post-treatment time series data were partitioned to improve learning of longitudinal tumor dynamics. The parameter encoder output for each patient was scaled by the corresponding time of last measurement to produce a set of kinetic rate parameters with the interpretation of inverse time as the physical unit (please note that the neural network schematics are only representational and do not reflect the actual layers or channel dimensionalities used). The initial condition ($\mathbf{z}(0)$) and kinetic parameters ($\mathbf{p}$) were then used in a neural-ODE model that represents the learned dynamical law and acts as a decoder of the system. Finally, the model solution was reduced to SLD predictions as a function of time. In parallel, the patients' OS is predicted using the ML model XGBSE. Via SHAP-ML and PCA analysis on the kinetic rate parameter distribution, our modeling paradigm successfully links tumor dynamics and OS in a data-driven manner.*

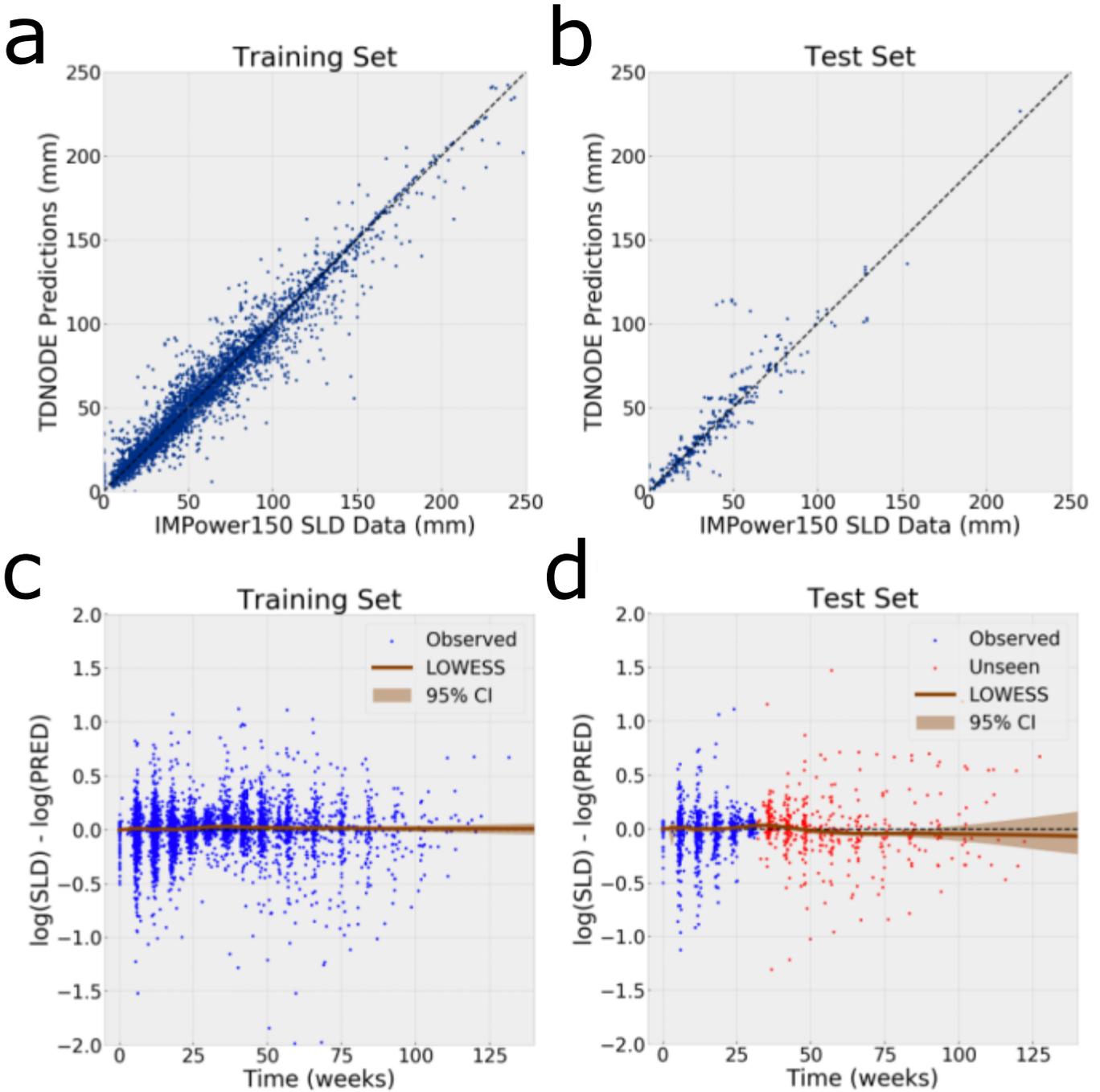

*Figure 2: TDNODE enables unbiased predictions of tumor dynamics with respect to both tumor size and observation time. (a) Prediction versus SLD data on the training set. (b) Prediction versus SLD data on the test set for t > 32 weeks. (c) Training set residuals between tumor dynamic predictions (PRED) and observed SLD data with respect to time. (d) Test set residuals between tumor dynamic predictions and observations with respect to time Bootstrapped LOWESS curves with 95% confidence intervals (CIs) were generated for c-d.*

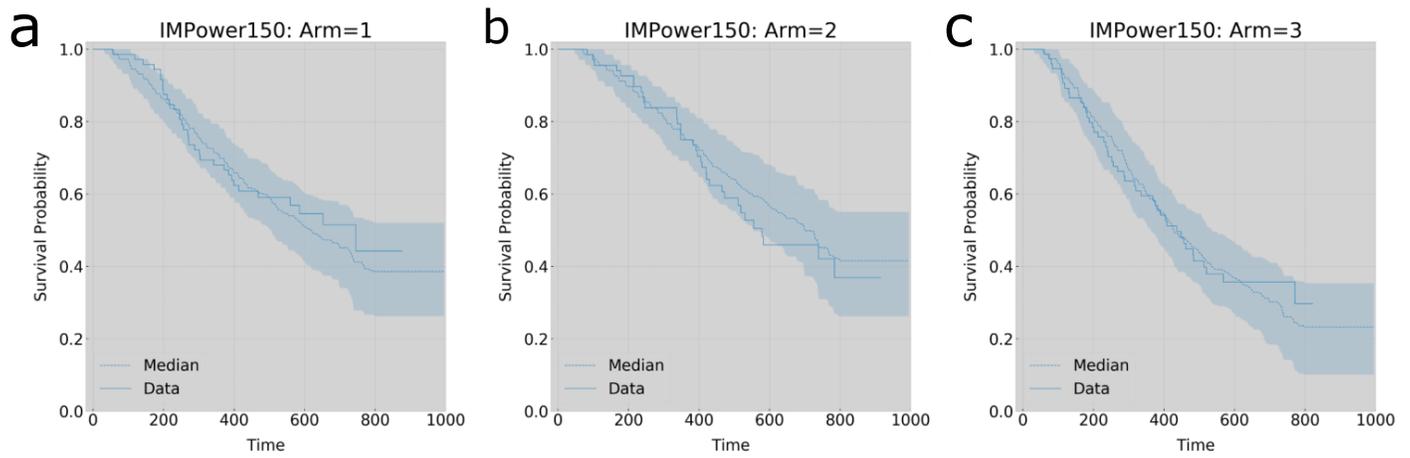

*Figure 3: Comparison of TDNODE prediction of survival probability with respect to treatment arm for test set patients (n = 216).* Predicted survival probability (median and 95% CI) using kinetic rate parameters versus data for patients enrolled in *(a)* Arm 1 (atezolizumab+ carboplatin + paclitaxel), n=73; *(b)* Arm 2 (atezolizumab+ carboplatin + paclitaxel + bevacizumab), n=68; *(c)* Arm 3 (carboplatin + paclitaxel + bevacizumab), n=75.

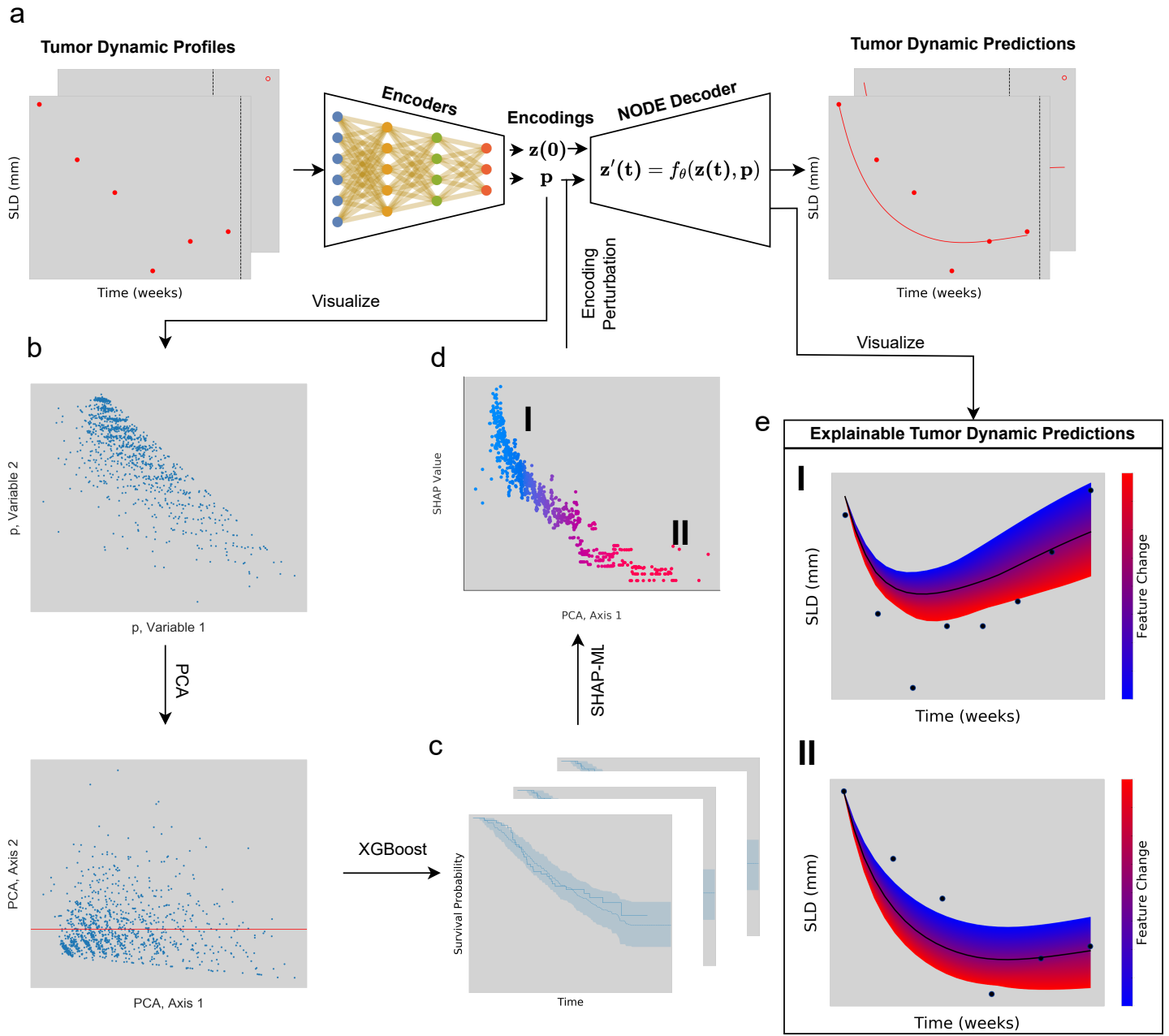

*Figure 4: Illustration of TDNODE derived metrics in the explainability of OS and tumor dynamic predictions for individual patients.* **(a)** Patient tumor dynamic data are fed into TDNODE encoders to produce 2-dimensional parameter encoding vectors that represent the patient's kinetic rate parameters, which are then fed into the NODE decoder to produce longitudinal tumor dynamic predictions. **(b)** We visualize the distribution of kinetic rate parameters for all patients in the dataset, prior to and after performing Principal Component Analysis (PCA). **(c)** Using only the first principal component, XGBoost is used to predict Overall Survival (OS). **(d)** SHAP analysis is performed to quantify the impact of the first principal component on OS; **I** and **II** refer to individual patients with differing PCA components and SHAP values. **(e)** Perturbations in longitudinal tumor dynamic profiles for each patient are made by systematically perturbing their encoder metrics along the direction of the first principal component, thereby linking tumor dynamic predictions to OS; charts **I** and **II** refer to the highlighted patients in **(d)**.

**Table 1: TDNODE predictive performance of tumor dynamics using a 32-week observation window on the test set.** *We evaluated the predictive performance of TDNODE on the unseen portion of the test set. For each patient, we let the observation window $w_i$ = 32 weeks and only evaluate measurements collected at time values beyond $w_i$. Although TDNODE generates a continuous solution of predictions $z(\cdot)$, the RMSE and $R^2$ scores are calculated using the discrete set of predictions only at observation times with SLD measurements. The predictive performance across all treatment arms is shown in bold. Variability was measured via median absolute deviation (MAD).*

| Treatment Arm | Number of Predictions for $t > w_i$ | RMSE (median$\pm$MAD) | $R^2$ Score (median$\pm$MAD) |
|---|---|---|---|
| Arm 1: atezolizumab+ carboplatin + paclitaxel | 208 | 12.56$\pm$1.29 | 0.75$\pm$0.05 |
| Arm 2: atezolizumab+ carboplatin + paclitaxel + bevacizumab | 214 | 7.65$\pm$0.93 | 0.93$\pm$0.01 |
| Arm 3: carboplatin + paclitaxel + bevacizumab | 79 | 4.34$\pm$0.31 | 0.98$\pm$0.01 |
| **All Treatment arms** | **501** | **9.69$\pm$0.75** | **0.88$\pm$0.02** |

**Table 2: Comparison of predictive performance for Overall Survival using TGI metrics and TDNODE derived metrics.** *Prediction of OS using TGI metrics compared with that of TDNODE encoder output metrics, both with and without eleven baseline covariates. The inclusion of these covariates significantly improved the prediction of OS in the TGI-OS model, whereas TDNODE metrics appear to capture the information provided by these covariates. In both cases, TDNODE generated metrics are superior for the prediction of patients' OS when compared to that of TGI-generated metrics. Variability of each metric is measured using median absolute deviation (MAD).*

| Model | Input features | C-index evaluated via 5-fold cross-validation (median±MAD) | C-index evaluated on the test set |
|---|---|---|---|
| TGI-OS.ML | TGI metrics only | 0.73±0.01 | 0.68 |
| TGI-OS.ML | TGI metrics + 11 baseline covariates | 0.78±0.02 | 0.75 |
| TDNODE-OS.ML | TDNODE metrics | **0.84±0.02** | **0.82** |
| TDNODE-OS.ML | TDNODE metrics + 11 baseline covariates | **0.86±0.02** | **0.84** |

# Supplementary Material:

# "Explainable Deep Learning for Tumor Dynamic Modeling and Overall Survival Prediction using Neural-ODE"

*Mark Laurie, James Lu*

## Supplementary Note 1

The model code available at:
- Github: https://github.com/jameslu01/TDNODE
- Zenodo: DOI 10.5281/zenodo.8436518

| Hyperparameter | Value |
|---|---|
| Batch Size | 8 |
| Learning Rate | 5.0e-5 |
| Optimization | ADAM |
| L2 Weight Decay | 1.0e-3 |
| Padding | Left-padding w/ first measurement |
| Epochs | 150 |
| Observation Window (weeks) | 32 |
| NODE Tolerance | 1.0e-4 |
| Tumor State Dimensionality ($c$) | 4 |
| Parameter Encoding dimensionality ($v$) | 2 |

**Supplementary Table 1: Select hyperparameter configurations for TDNODE.** *We selected and tuned hyperparameters based on empirically derived tumor dynamic predictivity and the OS c-index obtained from the XGBoost-ML model that was fitted with TDNODE-generated kinetic rate parameters. Seeking a model that could converge quickly, we also tuned hyperparameters, such as batch size, L2 Weight Decay, and NODE tolerance, considering the required optimization runtime.*

| Covariate | Description |
|---|---|
| CRP | C-reactive protein (mg/L) |
| BECOG | Baseline Eastern Cooperative Oncology Group (ECOG) Performance Status |
| LDH | Lactate Dehydrogenase (U/L) |
| NEU | Neutrophil Count ($10^9$/L) |
| METSITES | Number of Metastatic Sites at Enrollment |
| TPRO | Total Protein (g/L) |
| YSD | Number of Years Since Diagnosis (yr) |
| LIVER | Number of Liver Metastatic Sites at Enrollment |
| BNLR | Baseline Neutrophil to Lymphocyte Ratio |
| ALBU | Albumin (g/L) |
| HGB | Hemoglobin (g/L) |

***Supplementary Table 2: the set of baseline covariates used in the IMPower150 dataset.*** *The values of the baseline covariate were used within the XGBoost-ML models in addition to TDNODE-generated kinetic rate parameters.*

| Parameter | Description | Value |
|---|---|---|
| eta | Step size shrinkage | 0.0116 |
| max_depth | Maximum tree depth | 5 |
| min_child_weight | Child tree minimum required instance weight sum | 0.0211 |
| reg_alpha | L1 regularization term | 0.0014 |
| reg_lambda | L2 regularization term | 3.415 |
| subsample | Subsample ration to training instances | 0.849 |
| objective | Task specification | survival:cox |

***Supplementary Table 3***: ***The set of hyperparameters used in the XGBoost-ML models in predicting patients' OS.***

| Treatment Arm | Number of Measurements $t > w_i$ | RMSE (median±MAD) | $R^2$ Score (median±MAD) |
|---|---|---|---|
| Arm 1: Atezo.+Carb.+Pac. | 2,112 | 9.56±0.22 | 0.95±0.003 |
| Arm 2: Atezo.+Carb.+Pac.+Bev. | 2,453 | 8.72±0.18 | 0.94±0.003 |
| Arm 3: Carb.+Pac.+Bev. | 1,882 | 7.51±0.16 | 0.96±0.002 |
| **All Treatment arms** | **6,447** | **8.66±0.12** | **0.95±0.002** |

*Supplementary Table 4: Tumor dynamic predictivity of TDNODE measured via RMSE and $R^2$. For each patient, we let $w_i = 32$ weeks and only evaluate measurements collected beyond $w_i$. As displayed in the test set (**Table 1**), RMSE and $R^2$ scores are calculated using only the discrete set of predictions with corresponding observed SLD measurements. Variance was measured via median absolute deviation (MAD).*

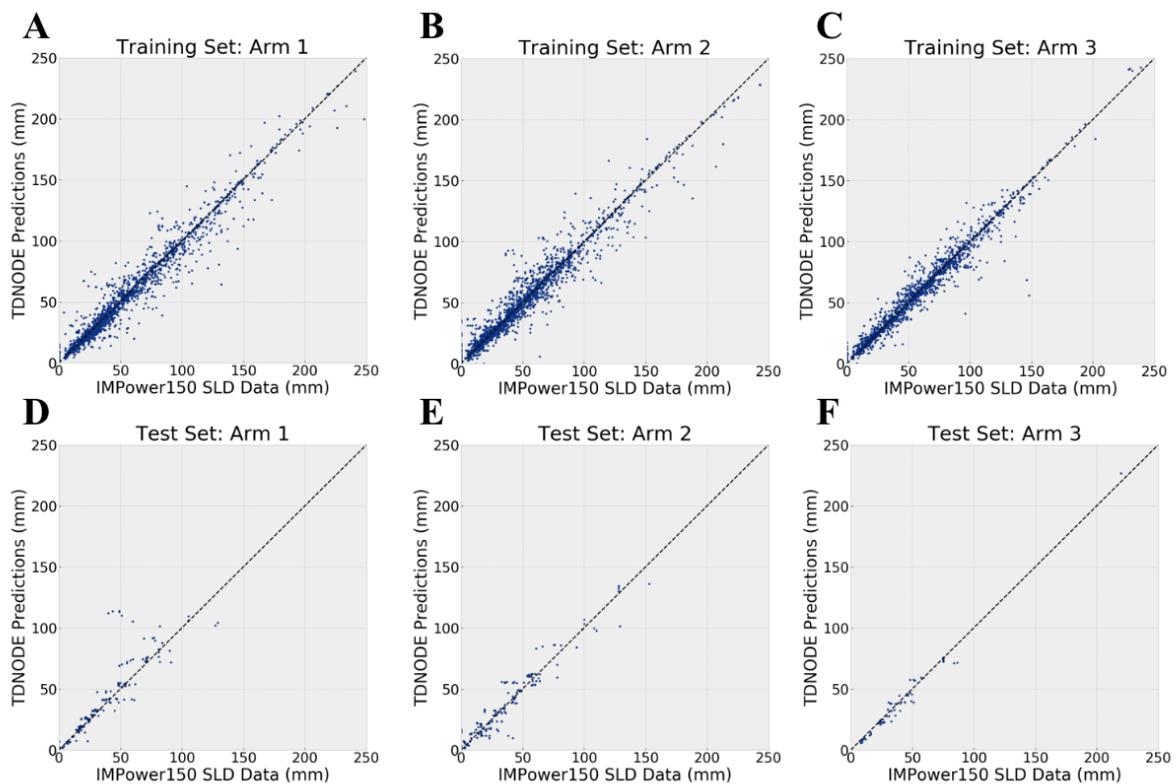

*Supplementary Figure 1: TDNODE enables unbiased tumor size predictions with respect to observed tumor size.*
***A-C**, Training set comparisons using all observed SLD data with respect to treatment arms 1, 2 and 3, respectively.*
***D-F**, Test set comparisons using only unseen SLD data with a 32-weeks observation window. Dashed lines denote the line of unity for each plot.*

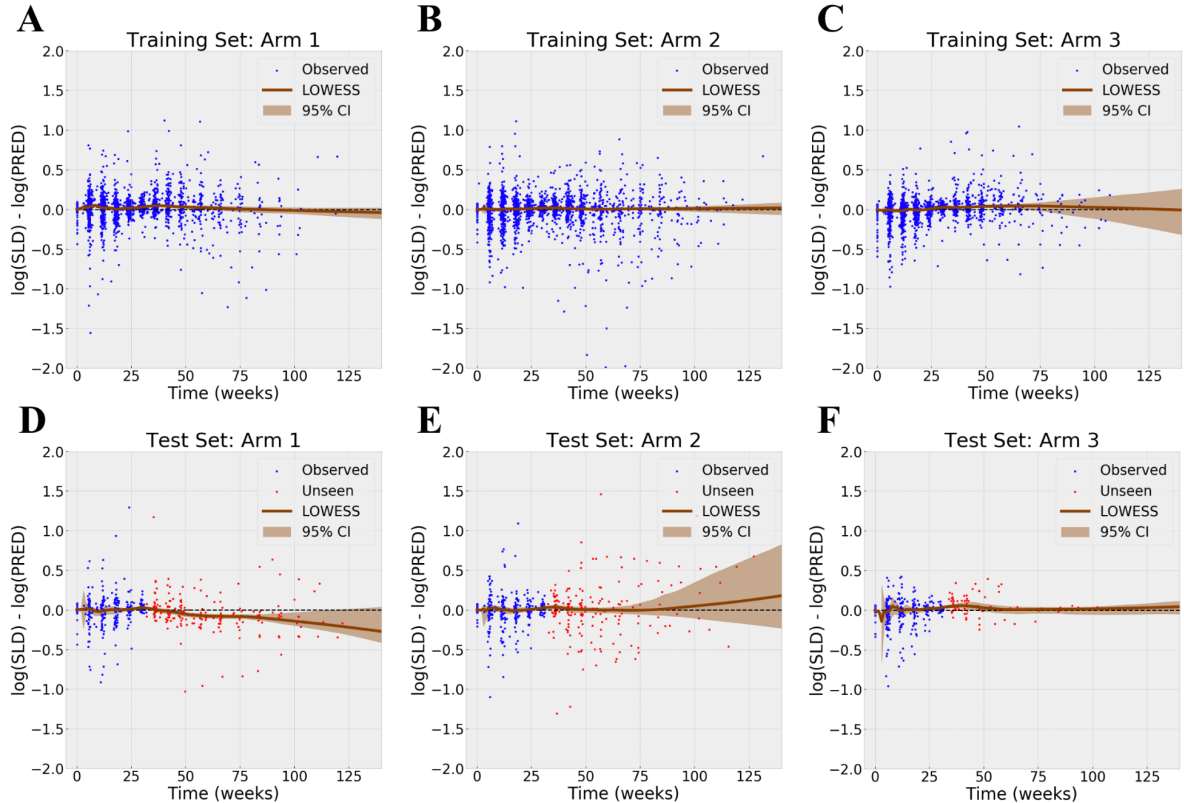

***Supplementary Figure 2: TDNODE enables unbiased predictions of tumor dynamics with respect to observation time. A-C.** Training set residuals using all observed SLD data with respect to time for treatment arms 1, 2 and 3, respectively. **D-F,** Test set residuals using all observed SLD data with respect to time for treatment arms 1, 2 and 3, respectively.*

| Observation Window (weeks) | Number of Predictions for $t > w_i$ | RMSE (median±MAD) | $R^2$ Score (median±MAD) |
|---|---|---|---|
| 16 | 928 | 13.48±0.47 | 0.82±0.02 |
| 20 | 768 | 12.22±0.51 | 0.83±0.02 |
| 24 | 682 | 10.52±0.51 | 0.87±0.02 |
| 28 | 616 | 10.89±0.59 | 0.86±0.02 |
| 32 | 501 | 9.69±1.02 | 0.88±0.03 |

***Supplementary Table 5: Tumor dynamic predictive performance of TDNODE upon variation of $w_i$, measured via RMSE and $R^2$.** Here we observe that TDNODE's predictive performance increases as the observations window for each patient is increased, as displayed by increasing RMSE and decreasing $R^2$ values. Variance is captured using median absolute deviation (MAD).*

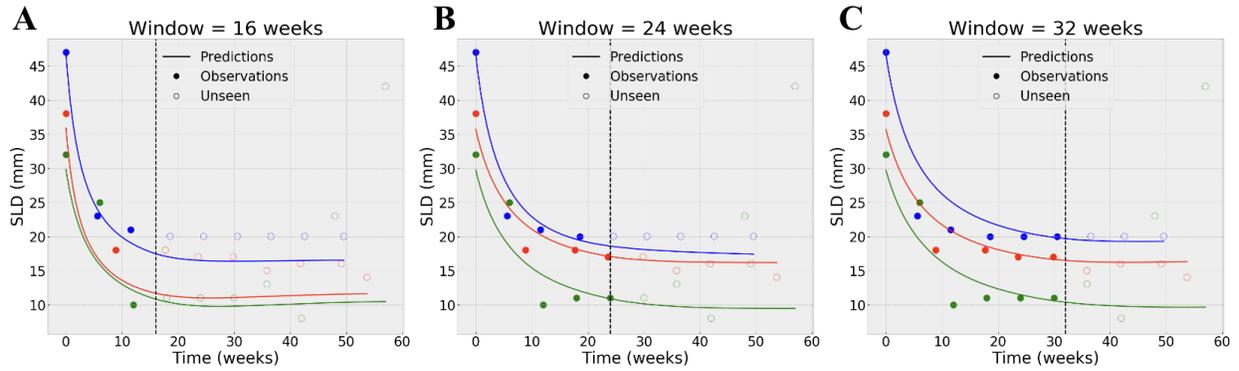

***Supplementary Figure 3: Illustration of tumor dynamic predictions using TDNODE with varying observation windows for selected test patients, showing increasingly more accurate predictions as the observation window is increased.** **A-C**, TDNODE tumor dynamic predictions for select test patients when $w_i$ = 16 weeks (**A**), $w_i$ = 24 weeks (**B**), $w_i$ = 32 weeks (**C**).*

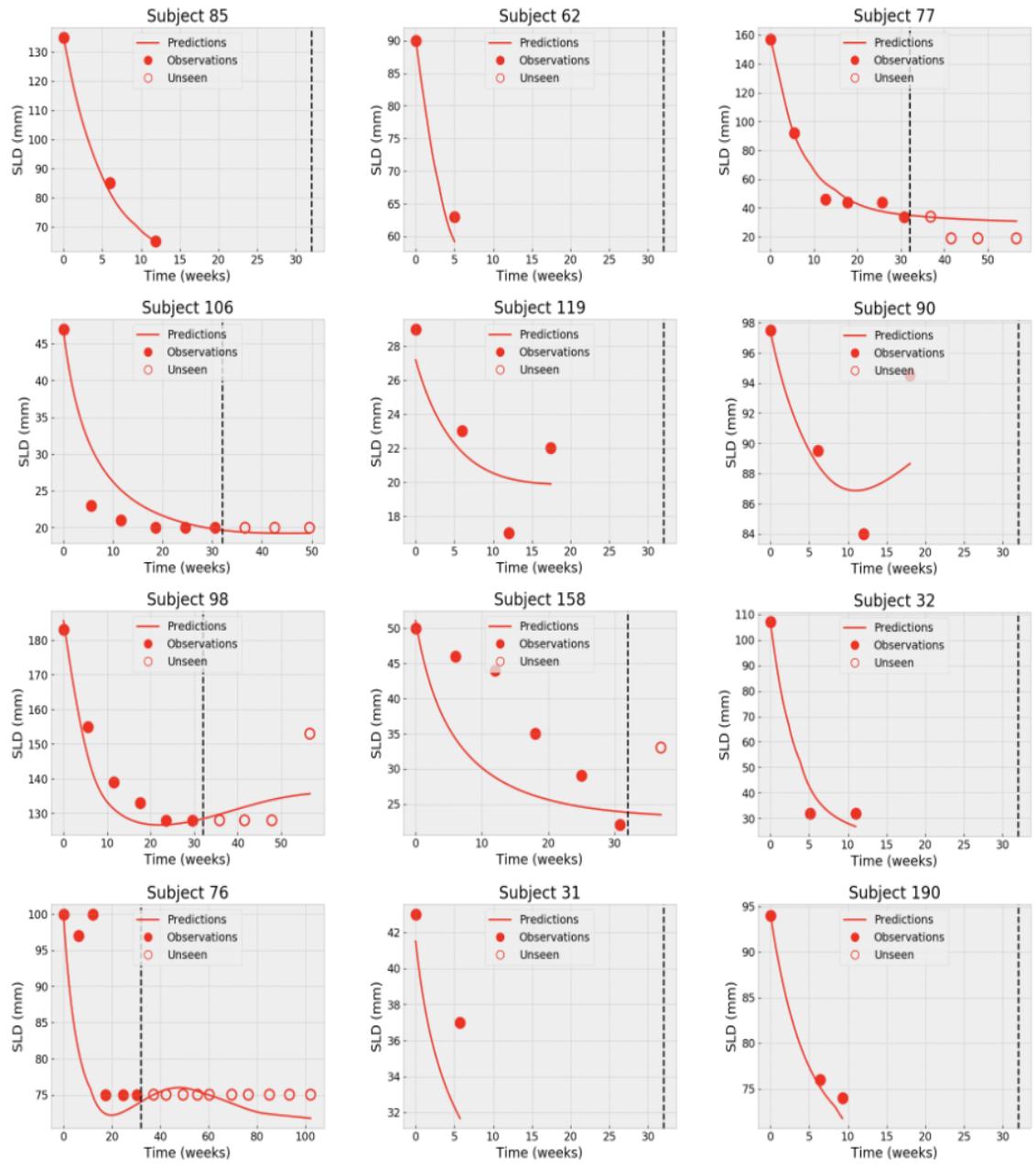

***Supplementary Figure 4: TDNODE individual prediction curves using a randomly selected sample of patients from the test set.*** *Here we observe how TDNODE performs on a set of patients with varying observed tumor dynamic profiles (e.g. remission, bimodal, failure) and quantity of observations in the test set with a 32-week observation window.*

| Input features | C-index evaluated via 5-fold cross-validation (median±STD) | C-index evaluated on test set |
|---|---|---|
| TDNODE metrics | 0.84±0.02 | 0.82 |
| TDNODE metrics + 11 baseline covariates | 0.86±0.02 | 0.84 |
| TDNODE Principal Components | 0.84±0.02 | 0.82 |
| TDNODE Principal Components + 11 baseline covariates | 0.85±0.02 | 0.84 |
| TDNODE Parameter Encoding 1st Principal Component | 0.83±0.02 | 0.81 |
| TDNODE Parameter Encoding 1st Principal Component + 11 baseline covariates | 0.84±0.02 | 0.83 |

***Supplementary Table 6: OS performance of 6 XGBoost-ML models trained with different sets of inputs.*** *Here we fitted 6 XGBoost-ML models with different sets of input data. We see that including baseline covariates do not significantly impact productivity of the OS. Using only the 1st principal component derived from the kinetic rate parameter distribution, we see that OS predictivity is comparable to that of using the original kinetic rate metrics.*

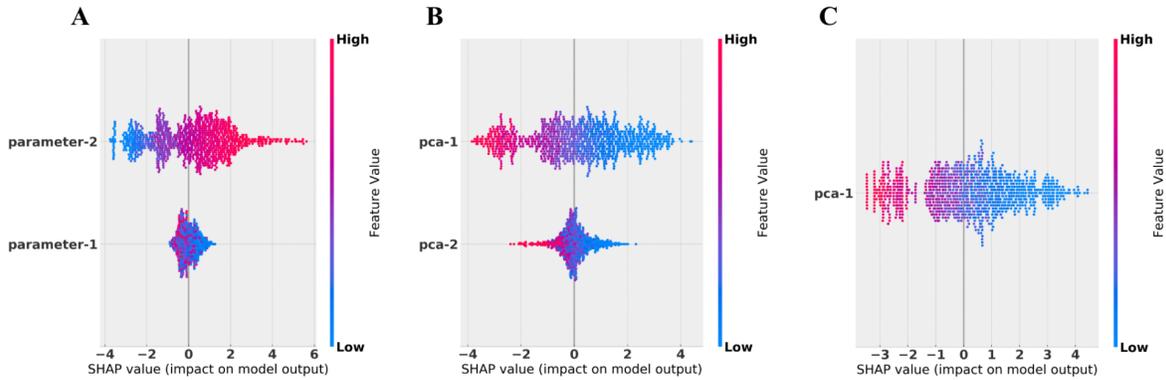

***Supplementary Figure 5: SHAP summary plot of XGBoost models utilizing TDNODE-derived metrics without baseline covariates***. ***A)*** *SHAP summary plot of XGBoost model using TDNODE-generated parameter encodings.* ***B)*** *SHAP summary plot of XGBoost model using the two most informative principal components of the encoding distribution.* ***C)*** *SHAP summary plot of XGBoost model using just the first principal component derived from the encoding distribution.*

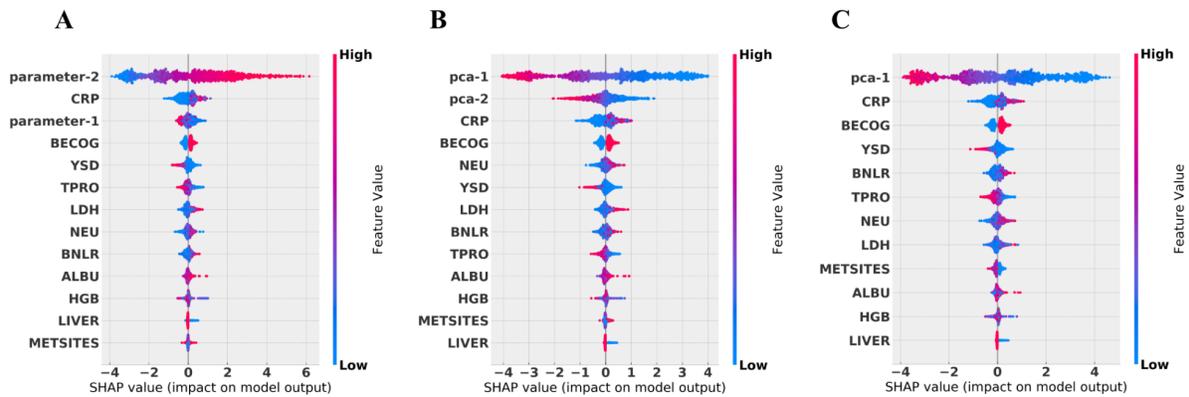

***Supplementary Figure 6: SHAP summary plot of XGBoost models utilizing TDNODE metrics with baseline covariates***. ***A) S****HAP summary plot of XGBoost model using TDNODE-generated parameter encodings.* ***B)*** *SHAP summary plot of XGBoost model using the two most informative principal components of the encoding distribution.* ***C)*** *SHAP summary plot of XGBoost model using just the first principal component derived from the encoding distribution.*

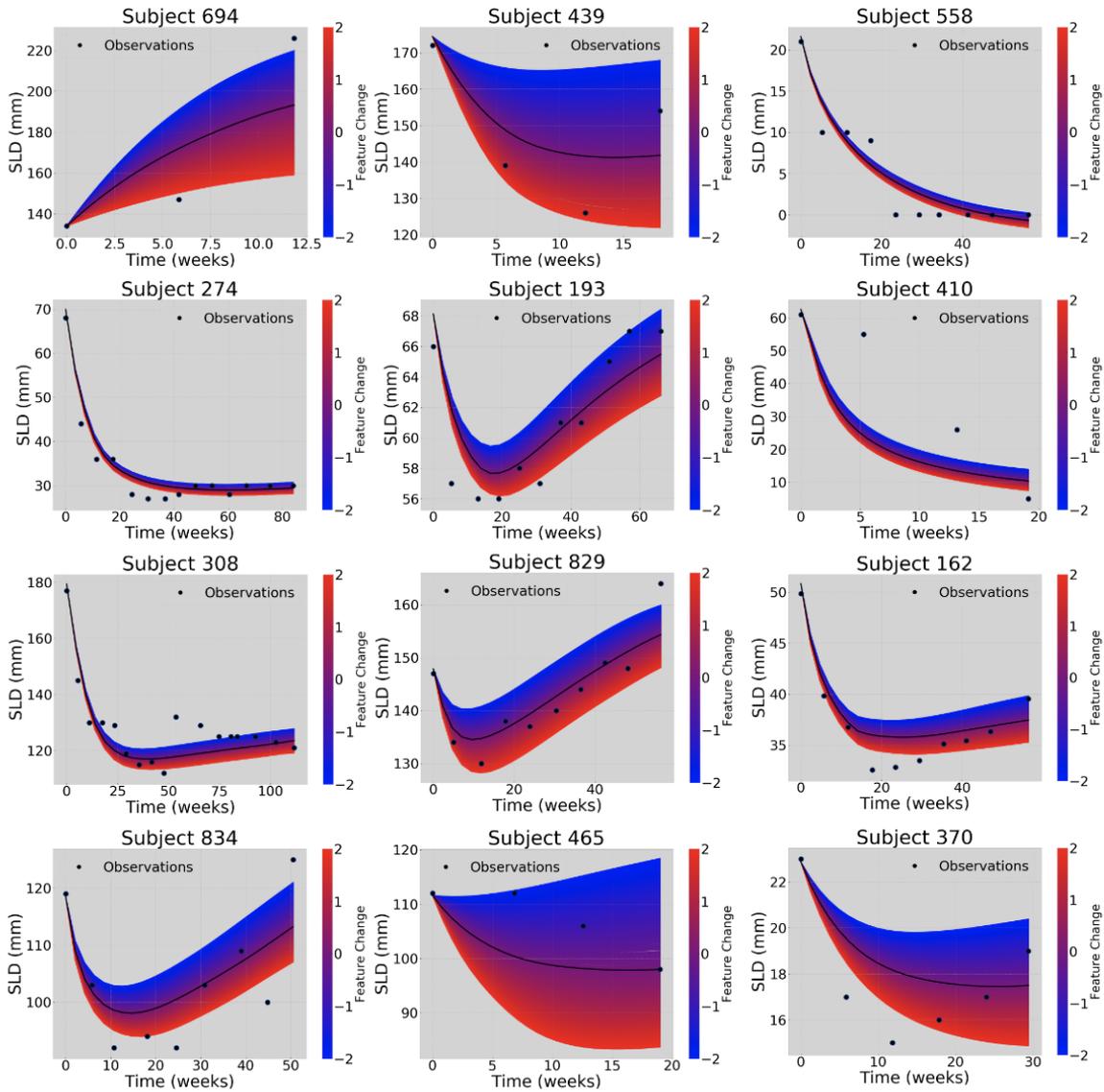

***Supplementary Figure 7: Additional feature dependence plots with perturbations of the first principal component calculated from the encoding distribution.*** *Color bar values correspond to the magnitude of change along the direction of the first principal component from that predicted by TDNODE, which is projected into the 2-dimensional parameter encoder space. Here we observe that increases in the value of the first principal component results in a decrease in the predicted tumor size. This finding corresponds with that derived from SHAP analysis of this principal component when used in XGBoost to predict OS, where increases in this principal component result in a negative change in corresponding patients' hazard rate.*

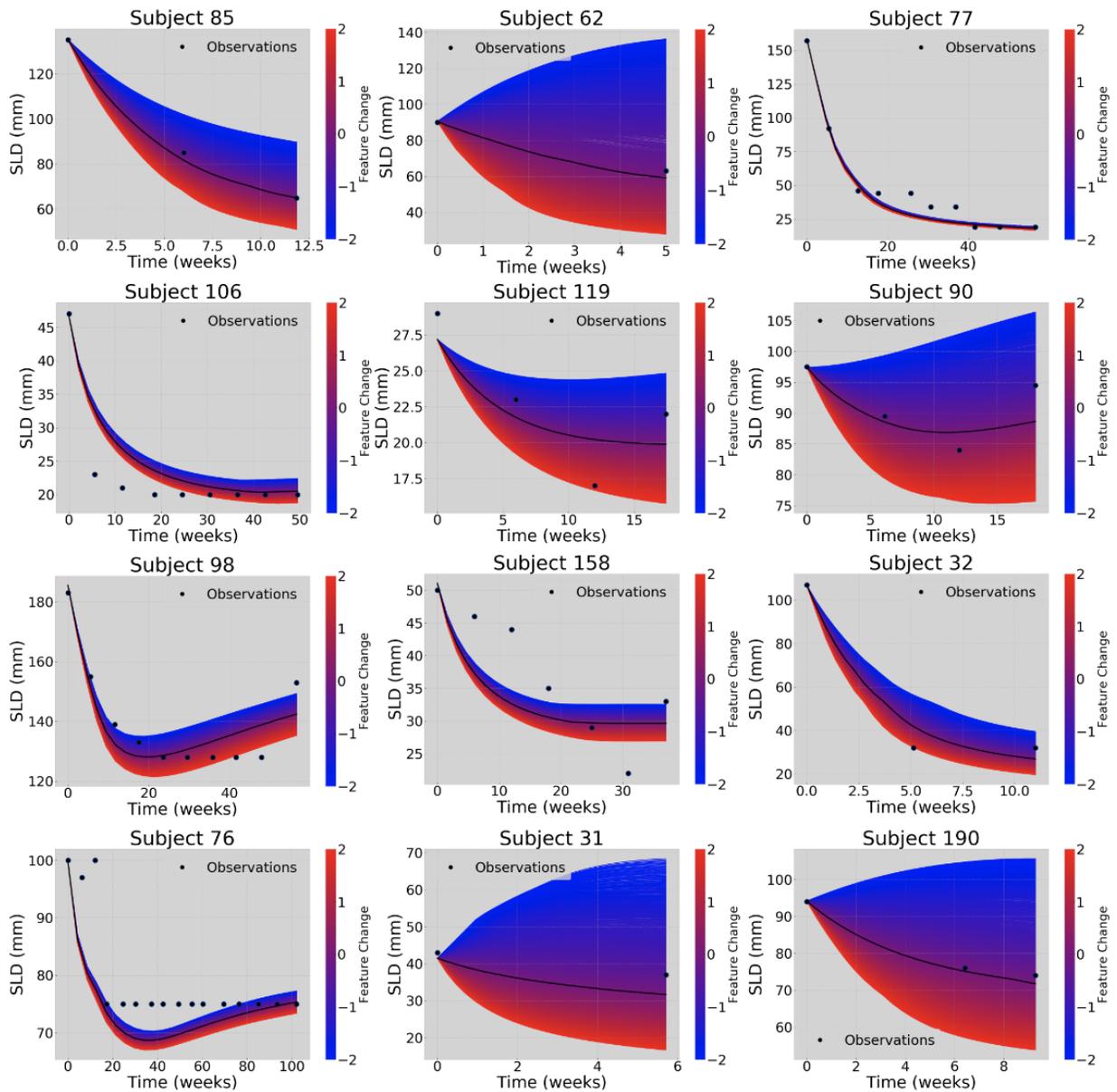

***Supplementary Figure 8: Additional feature dependence plots with systematic perturbations of kinetic parameters along the first principal component from their nominal values.*** *Color bar values correspond to the magnitude of change along the direction of the first principal component from that predicted by TDNODE, Changes in the 1st principal axis are projected into the 2-dimensional parameter encoder space. We use the same subjects as in **Supplementary Figure 4.** Like in the training set, we observe that increases in the value of the first principal component results in a decrease in the predicted tumor size. This finding corresponds with that derived from SHAP analysis of this principal component when used in XGBoost to predict OS, where increases in this principal component result in a net negative change in hazard rate.*

10